# A Radiative Cycle with Stimulated Emission from Atoms (Ions) in an Astrophysical Plasma.

S. Johansson[1] and V.S Letokhov[2,1]

[1]Lund Observatory, Lund University, P.O. Box 118, 221 00 Lund, Sweden

[2]Institute of Spectroscopy of Russian Academy of Sciences, 142190 Troitsk, Russia

We propose that a radiative cycle operates in atoms (ions) located in a rarefied gas in the vicinity of a hot star. Besides spontaneous transitions the cycle includes a stimulated transition in one very weak intermediate channel. This radiative "bottle neck" creates a population inversion, which for an appropriate column density results in amplification and stimulated radiation in the weak transition. The stimulated emission opens a fast decay channel leading to a fast radiative cycle in the atom (or ion). We apply this model by explaining two unusually bright Fe II lines at 250.7 and 250.9 nm in the UV spectrum of gas blobs close to η Carinae, one of the most massive and luminous stars in the Galaxy. The gas blobs are spatially resolved from the central star by the Hubble Space Telescope (HST). We also suggest that in the frame of a radiative cycle stimulated emission is a key phenomenon behind many spectral lines showing anomalous intensities in spectra of gas blobs outside eruptive stars.

Radiative processes play a key role in atomic astrophysics as stellar radiation controls the primary activity in the immediate surroundings of a star. Radiative mechanisms in low-density, non-equilibrium astrophysical plasmas manifest themselves in the spectrum by the appearance of unusually bright, sometimes time-variable, emission lines [1]. Enhanced lines are produced through "photoexcitation by accidental resonance" [2], in which an accidental (but not rare) wavelength coincidence between a strong line of the most abundant elements (H, He) and an absorption line from another element makes the photoexcitation possible. The most famous case is the Bowen mechanism [3]. However, in many cases such a coincidence cannot solely explain anomalous emission lines. For example, the resonant photoexcitation of an atom (ion) and the subsequent spontaneous decay may lead to an accumulation of atoms in an intermediate long-lived state, in which the atom resides for a long time ("a dark state"). Such a "bottle neck" in the spontaneous decay chain limits the cycling rate of photoexcitation + fluorescence, i.e. limits the efficiency of the conversion of UV photons into less energetic fluorescence from the absorbing atom. This effect is illustrated in Fig.1 in a simplified scheme of a 4-level atom, which represents a number of real cases in multilevel atoms (ions).

The "bottle neck" in the radiative decay chain, e.g. level 3 in Fig 1a, is fed by photoexcitation in $1 \rightarrow 4$ and a fast spontaneous decay, $4 \rightarrow 3$. This leads to an accumulation of atoms in the pseudo-metastable state 3 having a lifetime $\tau_3 \gg \tau_4$. For a high photoexcitation rate $W_{14}^{exc} \gg A_{32}$, the intensities of the fluorescent lines 4→3 and 2→1 are limited by the intensity of the "weak" transition 3→2 due to the transfer of atoms from the initial state 1 to the long-lived state 3. In a stellar atmosphere, for example, the density is high enough for collisional relaxation 3→2 to take place (Fig 1b). It is, however, effective in a cyclic process [4] only if state 2 is the ground state of the atom. Otherwise the collisions will open the closed cycle by relaxation from state 3 to other excited states. The goal of the present Letter is to demonstrate that in an ensemble of atoms under certain astrophysical conditions the radiative "bottle neck" leads to amplification and stimulated isotropic radiation in the spectral line $\lambda_{32}$. The accumulation of atoms in state 3 creates a population inversion in the transition 3→2 at an appropriate column density $\Delta N_{32}L$, where $\Delta N_{32}$ is the population density and $L$ is the size of the ensemble of atoms. If $W_{14}^{exc} \gg A_{32}$, the rate of stimulated transitions $W_{32}^{st} \gg A_{32}$, and the channel of weak spontaneous decay will be filled up by fast stimulated decay (Fig 1c). Thus, the rate for radiative cycling becomes enhanced as well as the energy transfer from the pumping UV line to the fluorescence lines. We show that such a radiative cycle with stimulated radiation operates in Fe II pumped by intense HLyα radiation in compact gas condensations (the so called Weigelt blobs [5]) close to η Carinae, one of the most luminous and massive stars in our Galaxy. This mechanism explains the origin of two unusually bright fluorescence lines at 250.7 and 250.9 nm observed in spectra of high spectral and spatial resolution recorded with the Hubble Space Telescope [6].

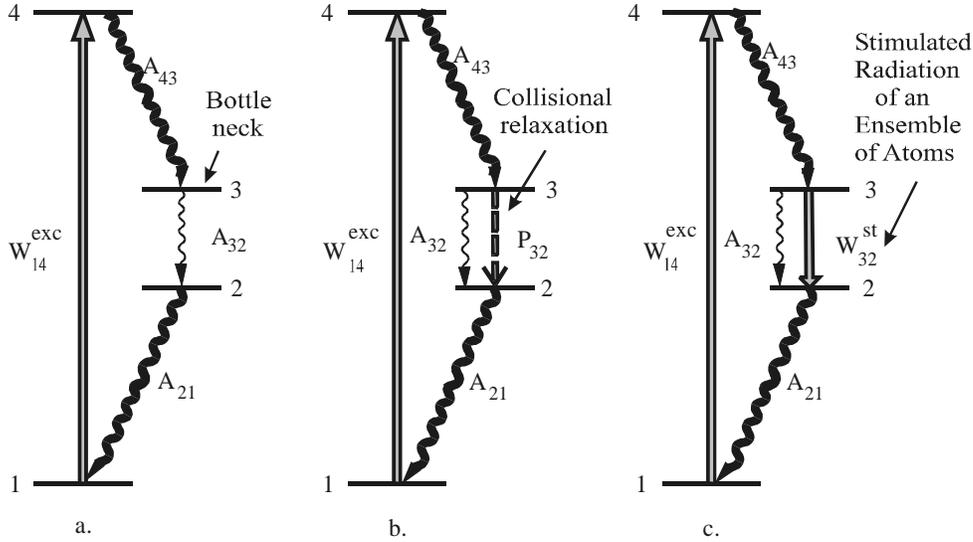

Fig. 1. A radiative cycle in a 4-level atom having allowed radiative transitions 1→4, 4→3, 2→1 and a metastable (pseudo-metastable) state 3 with a slow spontaneous radiative decay 3→2: a) An isolated atom without collisions (on the time scale $A_{32}^{-1}$), b) an atom in a high-density environment, which provides a fast collisional relaxation in the transition 3→2, c) a collision-free ensemble of atoms with an inverted population in the 3→2 transition and a significant amplification, which provides a fast stimulated radiative relaxation 3→2.

The diagram in Fig.2 includes a few of the known energy levels of Fe II, which together with the indicated transitions are relevant in the present study. The broad (a few hundreds cm$^{-1}$) HLyα line at 121.57 nm almost coincides in wavelength with two Fe II transitions from the low level $a^4D_{7/2}$ (level 1 in Fig. 2) to the close levels

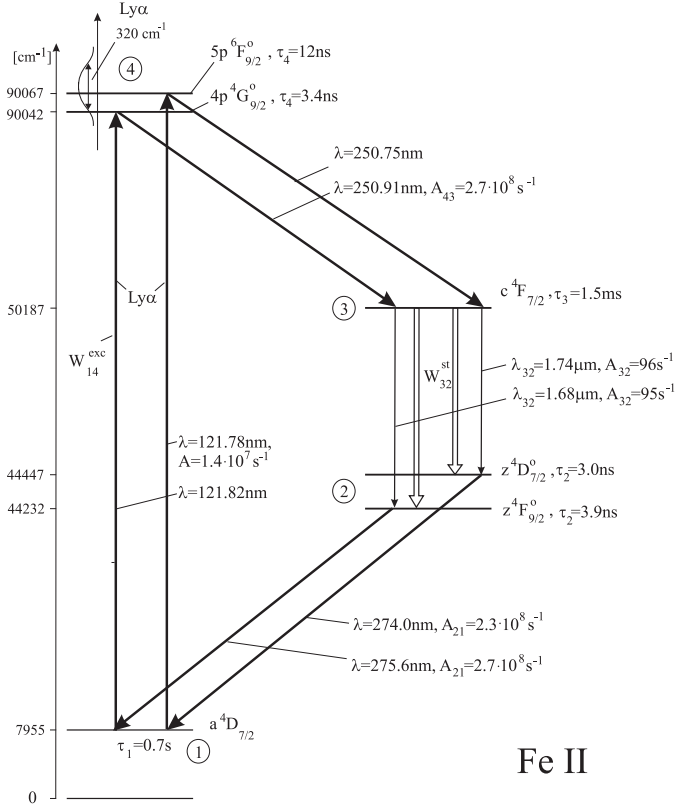

Fig. 2. A partial level diagram of the very complex multi-level ion Fe II forming an almost closed (on the time scale $\tau_1$) radiative cycle, which includes a selective photo-excitation of the uppermost level by intense H Ly$\alpha$ radiation. (References to the atomic data are given in [18].)

$(^5D)5p\,^6F^o_{9/2}$ and $(b^3F)4p\,^4G^o_{9/2}$ (marked 4 in Fig. 2), the frequency difference (detuning) being $\Delta\nu = -160$ cm$^{-1}$ and $-185$ cm$^{-1}$, respectively. The detuning is compensated by the broadening of HLy$\alpha$ when it penetrates the HI region having an optical depth of $\tau \geq 10^7 - 10^8$. This fluorescence channel was suggested and identified in the early work on IUE observations [7] and has later been confirmed by HST observations of the bright lines at 250.7 and 250.9 nm [6] corresponding to the transition 4→3 (see Fig. 2 for notations). The line at 250.91 nm to state 3 ($c\,^4F_{7/2}$) is the main decay channel from $(b^3F)4p\,^4G^o_{9/2}$ (state 4) with a branching fraction of about $\eta \approx 0.9$. Due to the small energy difference the two energy levels in state 4 ($(^5D)5p\,^6F^o_{9/2}$ and $(b^3F)4p\,^4G^o_{9/2}$) are strongly mixed and show the same decay pattern [8]. Level 3 is pseudo-metastable, i.e. it has a long lifetime (1.5 ms), but in contrast to metastable states it can decay slowly by electric dipole radiation to the short-lived states 2 ($z\,^4D^o_{7/2}$ and $z\,^4F^o_{9/2}$). It is essential that the decay of state 2 returns a large fraction of the Fe$^+$ ions to the initial state 1, which is metastable with a radiative lifetime of about 1 second. Let us emphasize, that we are considering a gas blob close to a star, the density (mainly hydrogen) is less than $10^{10}$ cm$^{-3}$ and the time scale for collisions is hence much longer than the time scale for radiative decay of state 3. Under these conditions the nearly closed radiative cycle 1→4→3→2→1 has a "bottle neck" in the 3→2 transition. If the rate of pumping $W_{14}^{exc}$ is larger than the decay rate

$1/\tau_3$ of state 3 ($\tau_3$=1.5 ms) an inverted population is created for the transition 3→2. The rate of selective photoexcitation of state 4 by isotropic HLyα radiation is defined by the expression

$$W_{14}^{exc} = A_{41} \left[ \exp\left(\frac{h\nu_{14}}{T_\alpha}\right) - 1 \right]^{-1}, \tag{1}$$

where the Einstein coefficient $A_{41}$=1.4·10$^7$ s$^{-1}$, and $T_\alpha$ is the brightness temperature of HLyα. Inversion of population ($W_{14}^{exc} > 1/\tau_3$) is achieved for $T_\alpha$>12000 K. The steady-state population inversion in the 3→2 transition is

$$\Delta N_{32} = N_3 - N_2 \approx N_3 = \frac{W_{14}^{exc} \tau_3}{1 + W_{14}^{exc} \tau_3} N_1, \tag{2}$$

where $N_2 \ll N_3$ because of the much faster decay of level 2, and $N_1$ is the concentration of Fe$^+$ in state 1 having a lifetime $\tau_1$=0.7s$\gg \tau_2, \tau_3, \tau_4$. At the brightness temperature $T_\alpha \approx$ (1.5-2.0)·10$^4$ K of HLyα in the Weigelt blob [9] the pumping rate $W_{14}^{exc} \gg 1/\tau_3$, leading to a full transfer of Fe$^+$ ions from state 1 to state 3. Using "laser" terminology this means that a strong saturation of the 3-level system 1→4→3 occurs due to the action of intense isotropic HLyα radiation.

The linear amplification coefficient for the 3→2 transitions at $\lambda_{32}$ = 1.74 μm and 1.68 μm (Fig. 2) is defined for $W_{14}^{exc} >$(2-3)/ $\tau_3$ by the standard expression

$$\alpha_{32} = \sigma_{32} \Delta N_{32} \approx \sigma_{32} N_3 \approx \sigma_{32} N_1. \tag{3}$$

The cross section $\sigma_{32}$ for stimulated emission of the 3→2 transitions in Fe II is

$$\sigma_{32} = \frac{\lambda_{32}^2}{2\pi} \cdot \frac{A_{32}}{2\pi \Delta \nu_D}, \tag{4}$$

where $\Delta \nu_D$ is the Doppler width. At a kinetic temperature of T=100-1000K in the relatively cold HI region (Fig. 3), $\Delta \nu_D \approx$ (200-600) MHz, i.e. $\sigma_{32} \approx$(1.2-3.7)·10$^{-16}$ cm$^2$.

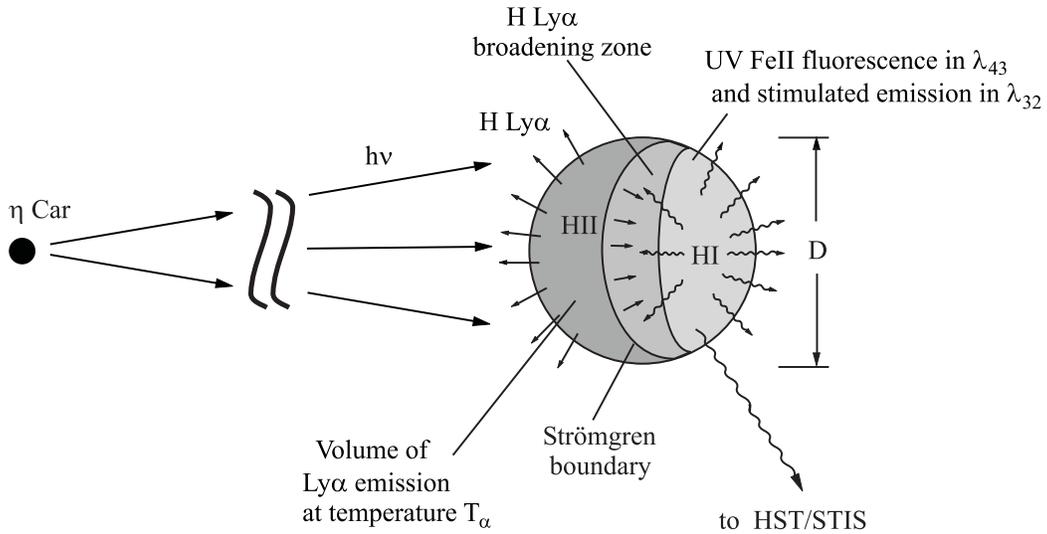

Fig. 3. An illustration of a physical model of a compact gas condensation in the vicinity of η Carinae showing the internal boundary between the ionized HII and neutral HI zones (the Strömgren boundary). Lyα radiation from the HII zone irradiates Fe$^+$ ions in the HI zone, resulting in a radiative cycle in Fe II with spontaneous as well as stimulated emission and a large enhancement of the spontaneous line $\lambda_{43}$.

The population in state 1, $N_1$, is determined by the fraction $f = N_1/N_0$, where $N_0$ is the total concentration of Fe$^+$ ions. The fraction $f$ is governed by the excitation rate of level 1 and its lifetime $\tau_1$. Most important for excitation of level 1 is the radiative decay of the high-lying states of Fe II excited by HLy$\alpha$ in about 10 transitions [7]. These decay channels can provide an excitation rate that is greater than $1/\tau_1 \approx 1.5$ s$^{-1}$ and, hence, sustain the relative population of state 1 at a level of $f \approx 0.01$. This would correspond to an approximately equal distribution of the Fe$^+$ ions among their 90 metastable and pseudo-metastable states, including state 1. Leaving the calculation of the magnitude of $f$ for future considerations, we adopt a qualitative estimate of $f \approx 0.01$.

The amplification coefficient for the 3→2 transition may, according to Eqn. 3, be estimated to $\alpha_{32} \approx (1.2\text{-}3.7) \cdot 10^{-18} \cdot N_0$ cm$^{-1}$, where $N_0 \approx 10^{-4} \cdot N_H$ is the density of iron, which is fully photoionized, and $N_H$ is the hydrogen density in the gas blob. According to data in [10] and calculations of the critical density of hydrogen [9], $N_H$ is estimated to be higher than $10^8$ cm$^{-3}$. Thus, for $\alpha_{32} \geq (1 - 4) \cdot 10^{-14}$ cm$^{-1}$ and a diameter D$\approx 10^{15}$ cm of the gas blob [10], which can be regarded as the length L of the amplifying region (Fig. 3), we get $\alpha_{32}L \approx (10\text{-}40)$ at a Ly$\alpha$ temperature $T_\alpha \geq 12000$ K, which corresponds to a large linear amplification coefficient $\exp(\alpha_{32}L)$. The large amplification coefficient $\alpha_{32}$ implies that stimulated emission occurs in the 3→2 transition (the $\lambda_{32}$ lines), and in the steady-state saturation regime the rate of stimulated transitions at 1.74 µm and 1.68 µm will be equal to the rate of photo-selective pumping $W_{14}^{exc}$ of state 3 ($W_{14}^{exc} \ll A_{43}$). The intensity (in photons/cm$^2$·s) of these isotropic laser lines should be enhanced by $W_{14}^{exc}/A_{32}$ relative to their intensity associated with spontaneous emission.

For an ensemble of atoms with sufficient density and size the channel 3→2 will open for fast stimulated decay, and the duration of the whole cycle is determined by the rate of the slowest excitation channel $W_{14}^{exc} = W_{32}^{st}$ rather than by the rate of the slowest spontaneous decay A$_{32}$. For a brightness temperature of Ly$\alpha$ in the range $T_\alpha \approx (1.5\text{-}2.0) \cdot 10^4$ K the excitation rate appears in the range $W_{14}^{exc} \approx 5 \cdot 10^3 \text{-} 2 \cdot 10^4$ s$^{-1}$ giving a duration of the total radiative cycle in Fe II of $\tau_{cycle} = 50\text{-}200$ µs. The Fe$^+$ ions can undergo a number of cycles ($\tau_{cycle}/\tau_1 \approx 3.5 \cdot 10^3 \text{-} 1.4 \cdot 10^4$) during the lifetime of state 1. A detailed study of the branching ratio effects and the anomalous intensity ratio of the strong UV lines and their satellites will be the subject of a forthcoming paper.

A large number of radiative cycles with a stimulated channel provide a large enhancement of the UV fluorescence lines (4→3) due to a suppressed accumulation of Fe$^+$ ions in the pseudo-metastable state 3 and the corresponding depletion of state 1. If the density $N_0$ and the size $L$ are not sufficiently large the amplification will be small and the stimulated channel will not operate. In such a case the radiative "bottle neck" limits the rate of the radiative cycling in Fe II to the time $\tau_1 = 1.5$ ms and the intensities of the two UV lines $\lambda_{43}$ become normal.

In fact, high-resolution spectra of the gas blobs outside $\eta$ Carinae, which are spatially resolved from the central star, show an anomalous brightness of the two UV lines [11]. A further observational evidence for the radiative cycle proposed in this Letter is the observation of the "bottle neck" lines around 1.7 µm in ground-based spectra of $\eta$ Carinae [12]. The 3→2 transition from c$^4$F$_{7/2}$ at $\lambda_{32} = 1.74$ µm and 1.68 µm appear relatively stronger than lines from the other fine structure levels of c$^4$F. Since the ground-based spectra contain integrated light from a larger region than the gas blob discussed here the absolute intensities cannot be compared with the intensities measured in the HST spectra. Moreover, according to the observations made in [11], the UV lines of the *2 → 1* transition at 274 and 275.6 nm have an integral intensity comparable with that of the bright *4 → 1* lines. Of course, the width of these lines is great because of the substantial optical density in the *4 → 1* transition and the corresponding resonance transfer radiation broadening. Resonance radiation trapping increases the effective lifetime of level *2*, but still it remains much shorter than the lifetime of level *3* and does not prevent the formation of an inverted population in the *3 → 2* transition.

The present Letter is related with the problem of astrophysical lasers, considered in [13] and recently discovered [14] in other Fe II transitions around 1μm in the same gas blob outside η Carinae as studied here. However, the new features in the present case compared to the case studied in [14] are the closed radiative cycle with stimulated emission transitions and the explanation of the two extremely bright UV lines. The astrophysical lasers discovered have no extreme brightness, but they transform very weak spectral lines of spontaneous emission to strong lines of stimulated isotropic radiation having intensities comparable to the intensity of allowed spectral lines. In the saturation regime the intensity of an astrophysical laser is limited by the rate of optical pumping, i.e. the intensity of the pumping Lyα line and the subsequent fast spontaneous decay to the upper laser level. There is an essential difference between an astrophysical laser [13,14] and an astrophysical maser [15-17]. The maser has an extraordinary brightness and effective temperature of microwave radiation due to the conversion of an unobservable pumping (without involvement of weak spontaneous microwave emission) into observable stimulated maser radiation. The effect of a cycle with both spontaneous and stimulated radiation proposed in this Letter enhances greatly the efficiency of optical excitation and, hence, the intensity of the optical lines involved in the cycle.

In conclusion, let us emphasize that stimulated emission at optical and near-infrared wavelengths forming non-collisional, radiative cycling of atomic particles in astrophysical media gives a natural explanation to the anomalous behaviour of spectral lines of two types: i) The transformation of a very weak line from a pseudo-metastable state having a low transition probability $A_{32}$ (line $\lambda_{32}$ in Fig. 2) into a strong line with an intensity comparable to that of an allowed transition and ii) the transformation of a normally intense line into an anomalously bright line (line $\lambda_{43}$ in Fig. 2) due to radiative cycling, which is only possible if stimulated emission drives the weakest link in the cycle.

We thank T. Gull and K. Ishibashi for providing reduced HST/STIS data. V.S.L. acknowledges financial support through a grant (S. J.) from the Royal Swedish Academy of Sciences and Lund Observatory for hospitality. The research project is supported by a grant (S. J.) from the Swedish National Space Board. The authors are grateful to the referees for very useful comments.

+-